\documentclass[10pt,aps,pra, superscriptaddress,floatfix,nofootinbib]{revtex4}
\usepackage{graphicx}
\usepackage[english]{babel}
\usepackage{amsmath}
\usepackage{amssymb}
\usepackage{ulem}
\usepackage[usenames,dvipsnames]{color}
\usepackage{color}
\definecolor{Blue}{rgb}{0.3,0.3,0.9}
\definecolor{orange}{rgb}{1,0.5,0}
\usepackage{subfigure}
\usepackage{bm}
\usepackage{titlesec}

\usepackage{mathrsfs}
\def\dbar{{\mathchar'26\mkern-12mu {\rm d}}}
\def\ddelta{{\mathchar'26\mkern-9mu {\rm \delta}}}

\newcommand{\inn}{\mbox{\scriptsize  in}}
\newcommand{\out}{\mbox{\scriptsize  out}}
\newcommand{\reg}{\mbox{\scriptsize  reg}}

\newcommand{\dd}{\mbox{d}}
\newcommand{\dS}{\mbox{d}{\mathbf\Sigma}}
\newcommand{\bx}{{\mathbf x}}
\newcommand{\by}{{\mathbf y}}
\newcommand{\bz}{{\mathbf z}}
\newcommand{\bk}{{\mathbf k}}
\newcommand{\bl}{{\mathbf l}}

\newcommand{\bv}{{\mathbf v}}
\newcommand{\sgn}{\mbox{sgn}}
\newcommand{\im}{\,{ \rm Im}\, }
\newcommand{\re}{\,{ \rm Re}\, }

\newcommand{\bbalpha}{{\bar\alpha}}

\begin{document}

\title{Quantum Cherenkov Radiation and Non-contact Friction}
\author{Mohammad F. Maghrebi}
\affiliation{Center for Theoretical Physics, Massachusetts Institute of Technology, Cambridge, MA 02139, USA}
\affiliation{Department of Physics, Massachusetts Institute of Technology, Cambridge, MA 02139, USA}
\author{Ramin Golestanian}
\affiliation{Rudolf Peierls Centre for Theoretical Physics, University of Oxford, Oxford OX1 3NP, UK}
\author{Mehran Kardar}
\affiliation{Department of Physics, Massachusetts Institute of Technology, Cambridge, MA 02139, USA}

\begin{abstract}
We present a number of arguments to demonstrate that a quantum analog of the Cherenkov effect occurs when two non-dispersive half-spaces are in relative motion. We show that they experience friction beyond a threshold velocity which, in their center-of-mass frame, is the phase speed of light within their medium, and the loss in mechanical energy is radiated  through the medium before getting fully absorbed in the form of heat. By deriving various correlation functions inside and outside the two half-spaces we explicitly compute this radiation, and discuss its dependence on the reference frame.
\end{abstract}

\maketitle

\section{Introduction}
An intriguing manifestation of quantum theory in macroscopic bodies is the non-contact friction between objects in relative motion. For example, two surfaces (or half-spaces) moving in parallel experience a frictional force if the objects' material is lossy~\cite{Pendry97,Volokitin99,Volokitin07}. The origin of this force is the quantum fluctuations of the electromagnetic field within and between the objects; the same fluctuations also give rise to Casimir/van der Waals forces. In brief, quantum fluctuations induce currents in each object, which then couple to result in the interaction between them. For moving objects,
a phase lag between currents leads to a frictional force between them.

For parallel plates (half-spaces), the friction force is related to the amplitude of the reflected wave upon scattering of an incident wave from each surface (formalized into a reflection matrix below) \cite{Pendry97,Volokitin99}.
Due to its quantum origin, friction persists even at zero temperature, where it is related to the imaginary part of the reflection matrix corresponding to evanescent waves.
It is usually assumed that the dielectric (or response) function itself has an imaginary part due to dissipative properties of the material; this then leads to an imaginary reflection matrix and hence friction \cite{Dalvit11}.
However, this is not necessary as, even for a vanishingly small loss, evanescent waves lead to an imaginary reflection matrix.

We consider non-dispersive half-spaces described by a real constant dielectric function, such that light propagates in the medium with a constant (reduced) speed. Note that the frequency-independence of the dielectric function follows from a vanishing imaginary part, due to Kramers-Kronig relations.
We show that when the velocity of moving half-spaces, in their \emph{center-of-mass} frame, is larger than the phase speed of light in the medium, a frictional force arises between them. This is in fact a quantum analog of the well-known  \emph{classical Cherenkov} radiation.
We elaborate on the relation between the friction, and radiation in the gap as well as within the half-spaces. We emphasize, however, that {\it dispersive} half-spaces can experience friction at any velocity. Even non-dispersive bodies moving non-uniformly experience vacuum friction at arbitrarily low speeds.

Quantum Cherenkov radiation was first discovered by Ginzburg and Frank~\cite{Frank45} in a rather different setup. They argued that when an object (an atom, for example) moves inertially and superluminally, i.e. larger than the phase speed of light in a medium, it spontaneously emits photons; see Refs.~\cite{Ginzburg93,Ginzburg96} for subsequent reviews by Ginzburg. This phenomenon is intimately related to \emph{superradiance}, first discovered by Zel'dovich~\cite{Zel'dovich71} in the context of rotating objects and black holes: A rotating body amplifies certain incident waves even if it is lossy. The underlying physics is that a moving object (atom) can lose energy by getting excited. This is because, at superluminal velocities, an excitation in the rest-frame of the object corresponds to a loss of energy in the \emph{lab} frame. Ginzburg and Frank refer to this eventuality as the \emph{anomalous} Doppler effect~\cite{Frank45}; see also Ref.~\cite{Bekenstein98}.

Since these unusual observations span several subfields of physics, we find it useful to demonstrate the results by a number of different formalisms. We first generalize the arguments by Ginzburg and Frank to prove dissipation effects associated with the relative motion of two parallel plates. We then use the input--output formalism of quantum optics to derive and compute the friction force based on  scattering matrices. An alternative proof follows approaches introduced in the context of quantum field theory in curved space-time, making use of an inner product to identify the wavefunctions and their (quantum) character.
Application of the latter formalism to vacuum friction
is particularly suited to a real dielectric function, and to the best of our knowledge
has not been employed in this context before.
Finally we employ the Rytov formalism~\cite{Rytov89} which is grounded in the fluctuation-dissipation theorem for electrodynamics and well-known to practitioners of non-contact friction. We thereby extend previous results on friction, and radiation in the gap between the half-spaces to those within their medium, which is desired to establish a connection to the Cherenkov radiation.

To ease computations, however, we consider a scalar field theory as a simpler substitute for electromagnetism. The former shares the same conceptual complexity while being more tractable analytically. This is particularly useful in expressing complicated Green's functions with points both inside and outside each half-space, or within the gap between them. The generalization to vector and dyadic electromagnetic expressions should be straightforward but laborious.
Finally to avoid complications of the full Lorentz transformations, we limit ourselves to small velocities---both the relative velocity of the objects and the speed of light in their media.

The remainder of the paper is organized as follows. In Sec.~\ref{Sec: Friction}, existing formulas are used without proof to compute friction, and to discuss the similarities with the classical Cherenkov effect. In this section, we elaborate on friction in a specific example. In Sec.~\ref{Sec: Arguments}, we consider a general setup, argue for and derive the friction force, as well as emitted radiation, in great detail. This section comprises 4 subsections each devoted to one particular formalism. Specifically, we discuss how the radiation within the half-spaces, and in the gap, depend on the reference frame.

\section{Friction}\label{Sec: Friction}
 We start with a scalar model that is described by a free field theory in empty space, while inside the medium a ``dielectric'' (or, a response) function $\epsilon$ is assumed which characterizes  the object's dispersive properties. The field equation for this model reads
  \begin{equation}\label{Eq. field equation}
  \left(\nabla^2+\epsilon(\omega,\bx) \frac{\omega^2}{c^2}  \right)\Phi(\omega,\bx)=0,
  \end{equation}
with $\epsilon=1$ in the vacuum, and a frequency-dependent constant inside the medium.

We consider the configuration of two parallel half-spaces
in $D$ spatial dimensions, separated by a vacuum gap of size $d$.
For each half-space in its rest frame, a plane wave of frequency $\omega$ and wavevector
$\bk$ is reflected with amplitude (`reflection matrix')
  \begin{align}\label{Eq: reflection matrix}
    R_{\omega \bk_\|}=-\frac{\sqrt{\epsilon \, \omega^2/c^2-\bk_\|^2}-\sqrt{\omega^2/c^2-\bk_\|^2}}{\sqrt{\epsilon \, \omega^2/c^2-\bk_\|^2}+\sqrt{\omega^2/c^2-\bk_\|^2}},
  \end{align}
where $\bk_\|$ is the component of the wavevector parallel to the surface. This result is easily obtained by solving the field equations inside and outside the half-space, and matching the reflection amplitude to satisfy the continuity of the field and its first derivative along the boundary.
We are particularly interested in friction at zero temperature which is mediated solely by evanescent waves~\cite{Pendry97,Volokitin99,Volokitin07}; for further discussion see Ref.~\cite{dedkov10}. Such waves contribute to friction through the imaginary part of the reflection matrices.
If one half-space moves  laterally with velocity $v$ along the $x$ axis, while the other is at rest,
the friction force is given by (introducing the notation $\dbar x={\dd x}/{2\pi}$)
  \begin{align}\label{Eq: Friction}
    f=&\int_{0}^{\infty} \dbar \omega \, L^{D-1}\!\int\dbar \bk_\| \,\, \hbar k_x
    \frac{e^{-2 |k_\perp| d}  \, (2 \im R_1) \,(2\im R_2)}{{|1-e^{-2|k_\perp| d} R_1 R_2|^2}} \,\Theta(-\omega+ v k_x)\, ,
  \end{align}
where $\Theta$ is the Heaviside step function, $k_\perp=\sqrt{\omega^2/c^2-\bk_\|^2}$, and $L^{D-1}$ is the area. Note that the reflection matrix of the static half-space is given by Eq.~(\ref{Eq: reflection matrix}), but that of the moving half-space is obtained after Lorentz transforming to the lab frame.

We leave the derivation and extension of Eq.~(\ref{Eq: Friction}) to the next section, but discuss its implications here.
While this equation has been studied extensively in the literature, it is usually assumed that the dielectric medium is lossy, with a nonzero imaginary part of $\epsilon$.
However, even when $\im \epsilon$ is vanishingly small, a frictional force can be obtained as follows.
With $\im \epsilon\approx0$, the medium can be characterized by the modified speed of light $v_0={c}/{\sqrt{\epsilon}}$. The only relevant length scale in the problem (aside from the
overall area $L^{D-1}$) is the separation  $d$.
We can then construct the frictional force on purely dimensional grounds as
\begin{align}
\nonumber
    f= \frac{\hbar \, v_0 L^{D-1}}{d^{D+2}} \, \tilde g\left(\frac{v}{v_0}, \frac{v}{c}\right)\,,
  \end{align}
where $\tilde g$ is a function of two dimensionless velocity ratios.
Any velocity could have appeared as prefactor (with a correspondingly modified function $\tilde g$);
we have chosen $v_0$ for convenience.
For small velocities, the dependence on vacuum light velocity $c$ drops out\footnote{One can see this explicitly from Eq.~(\ref{Eq: Friction}).}, and
  \begin{align}
    f\approx \frac{\hbar \, v_0 L^{D-1}}{d^{D+2}} \, g\left(\frac{v}{v_0}\right)\,,
  \end{align}
  with $g$  depending only on the ratio of the velocity $v$ to the light speed in the medium $v_0$. Interestingly, at small $v$, only the modified speed within the media is relevant. Our assumption pertains to the non-retarded limit when the speed of light can be formally taken to $c\to \infty$. Barton has also considered the same limit in Ref.~\cite{barton11} where he computes the frictional (drag) force between weakly dissipative media described by the {\it Drude model}, hence
  obtaining different power laws  in the limit of zero temperature.

  Now note that the Heaviside function in Eq.~(\ref{Eq: Friction}) restricts to frequencies
  \begin{equation}
    (0<)\,\omega <v k_x\, .
  \end{equation}
  Furthermore, the imaginary part of the reflection matrix $R_1$, given by Eq.~(\ref{Eq: reflection matrix}), is only nonzero when $\omega^2/c^2-\bk_\|^2<0$ {and} $\epsilon\,\omega^2/c^2-\bk_\|^2>0$, which, in turn, imply
  \begin{equation}\label{Eq: first plate}
    |\omega|> v_0 |\bk_\||>v_0 |k_x|.
  \end{equation}
A similar condition holds for the second half-space: $|\omega'|>v_0 |k'_x|$ with primed values defined in the moving reference-frame.
For simplicity, we assume that $v, v_0 \ll c$, and thus neglect the complications of a full Lorentz transformation. Hence, $\omega'\approx\omega-v k_x$ and $k_x'\approx k_x- v \omega/c^2\approx k_x$.
Then the analog of Eq.~(\ref{Eq: first plate}) for the second half-space reads
  \begin{equation}
    |\omega-v k_x|\gtrapprox v_0 |k_x|.
  \end{equation}
  \begin{figure}[h]
    \centering
  \includegraphics[width=12cm]{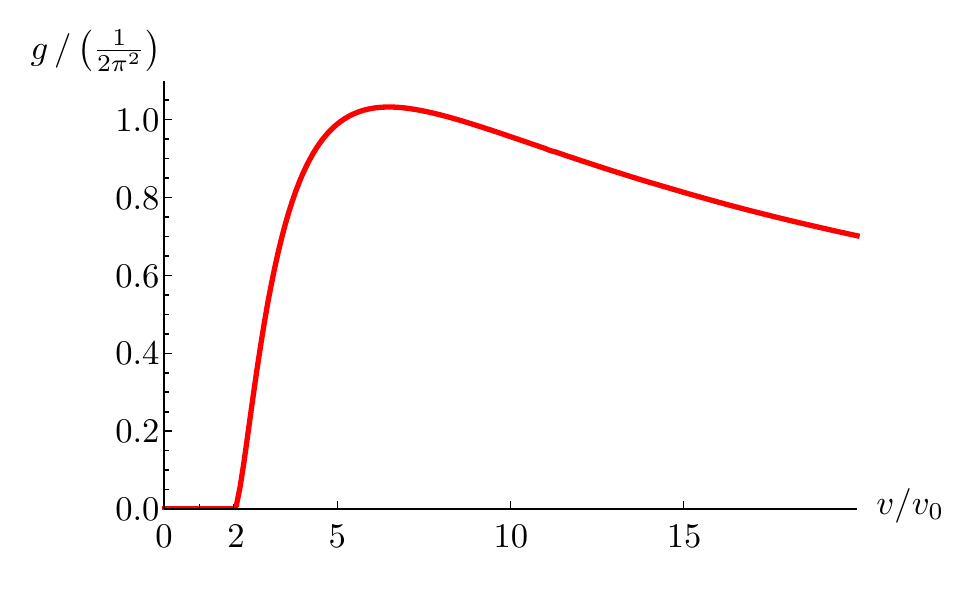}
   \caption{Friction depends on velocity $v$ through the function $g$. Below a certain velocity, $v_{\rm min}=2v_0=2{c}/{\sqrt{\epsilon}}$, the friction force is zero; it starts to rise linearly at $v_{\rm min}$, achieves a maximum and then  falls off.} \label{Plot: Cherenkov}
 \end{figure}
  The above conditions limit the range of integration to
  \begin{align}
    k_x>0, \qquad \mbox{ and} \qquad v_0 k_x <\omega <(v-v_0)k_x.
  \end{align}
  One then finds the minimum velocity where a frictional force arises as
  \begin{equation}\label{Eq: v-min}
    v_{\rm min}= 2 v_0=2 \frac{c}{\sqrt{\epsilon}}.
  \end{equation}
  This threshold velocity is reminiscent of the classical Cherenkov effect, although larger by a factor of two. However, in the \emph{center-of-mass} frame where the two half-spaces move at the same velocity but in opposite directions, we find the same condition as that of the Cherenkov effect: A frictional force arises when, in the   center-of-mass frame, the half-spaces' velocity exceeds that of  light in the medium.
  As a specific example, we consider a two dimensional space, i.e.  surfaces represented by straight lines. The dependence of the friction force on relative velocity is then plotted in Fig.~\ref{Plot: Cherenkov}.

\section{Formalism and Derivation}\label{Sec: Arguments}
In the previous section we argued for the appearance of friction between moving parallel plates which is reminiscent of the Cherenkov effect.
Establishing a complete correspondence requires a full analysis of the radiation within each object.  In this section, we provide several arguments to demonstrate why and how a fluctuation-induced friction arises in the context of macroscopic objects in relative motion. Our presentation is not a repetition of the existing literature:
Friction, as well as radiation within the gap, are obtained through novel
methodologies, and further extended to compute radiation inside the media.
We start with a heuristic argument, making a connection with the Frank-Ginzburg condition. We then present three distinct derivations of the friction force using techniques developed in different fields: The first method relies on the input-output formalism in a second-quantized picture; the second one appeals to quantum field theory in curved space-time. The last two approaches are, to the best of our knowledge, novel in their application to quantum friction between moving half-spaces. The last, and the longest, derivation is based on fluctuation-dissipation theorem, or the closely related Rytov formalism. The advantage of the latter approach is in finding correlation functions inside and outside the two half-spaces which can be used to compute the radiation within each half-space and in the gap between them.

\subsection{Why is there any friction/radiation?}\label{Sec: Frank-Ginzburg}
  To start with, let us consider a space-filling dielectric medium described by a constant real $\epsilon$. A wave described by  wavevector $\bk$ satisfies the dispersion relation $\omega=v_0 |\bk|$, with $v_0={c}/{\sqrt{\epsilon}}$ being the speed of light in the medium. This relation describes the spectrum of quantum field excitations. If the medium is set in motion, the new spectrum can be deduced simply by a Lorentz transformation from the static to the moving frame.
Assuming again  that the speed of light in medium is small (or $\epsilon$ is large), we find
  \begin{equation}\label{Eq: spectrum}
    \omega= v_0 |\bk|+v k_x\,,
  \end{equation}
with the medium moving with speed $v$ parallel to the $x$ axis.

 \begin{figure}[h]
    \centering
  \includegraphics[width=13cm]{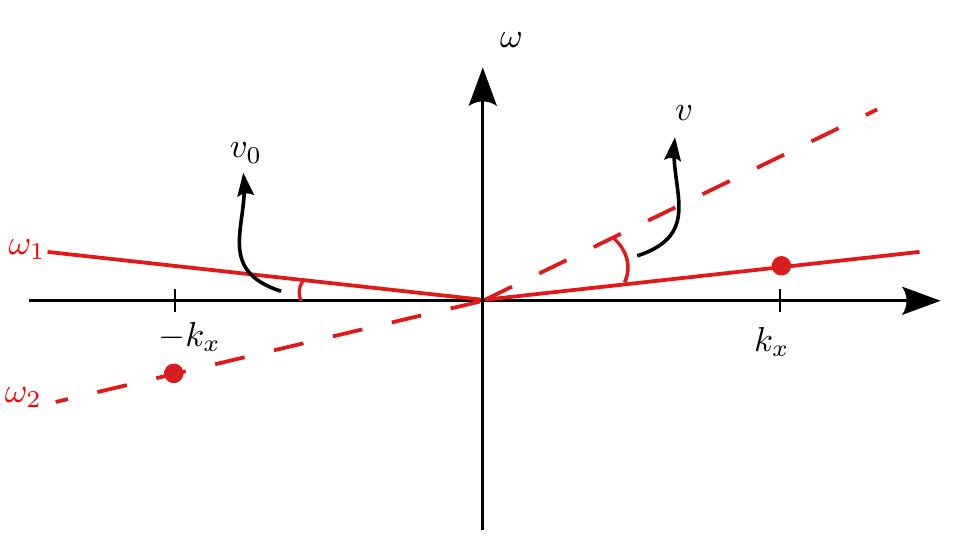}
   \caption{The energy spectra for a medium at rest (solid curve denoted by $\omega_1$), and a moving medium (dashed curve denoted by $\omega_2$),
in the $(\omega,k_x)$ plane. The spectrum for the moving medium is merely tilted.
The production of a pair of excitations, indicated by solid circles at opposite momenta, is energetically possible for $v>2 v_0$.} \label{Fig: Spectrum}
 \end{figure}

Next  consider two (semi-infinite) media one of which moves laterally with velocity $v$, whereas the other is at rest. Although the boundaries modify the dispersion relation, we may assume that Eq.~(\ref{Eq: spectrum}) {\it approximately} describes each medium (with $v=0$ for the stationary body).
This is  justified by considering wavepackets away from boundaries. We thus have two distinct spectra: The spectrum for the half-space at rest is akin to a cone while that of the moving half-space is \emph{tilted} towards the positive $x$ axis, as in Fig.~\ref{Fig: Spectrum}. For the sake of simplicity, we limit ourselves to $\bk=k_x \hat {\bf x}$. Let us consider the (spontaneous) production of two particles, one in each medium. Since linear momentum is conserved the two particles must have opposite momenta ($k_x$ and $-k_x$). This process is energetically favored if the sum of the energy of the two particles is negative, that is, spontaneous pair production occurs if it lowers the energy of the composite system. This condition is satisfied when
  \begin{equation}\label{Eq: energetics}
    \omega_1+\omega_2=2 v_0 |k_x| +v k_x <0\,,
  \end{equation}
  which is possible only if $v> 2 v_0$. We stress that our argument is not specific to a particular reference frame. If both half-spaces are moving with velocities $v_1$ and $v_2$, the velocity $v$ in Eq.~(\ref{Eq: energetics}) is replaced by $v_2-v_1$, thus this equation puts a bound on the relative velocity.

The above argument is similar to the Landau criterion for obtaining the critical velocity of a superfluid flowing past a wall~\cite{Landau41}. The instability of the quantum state against spontaneous production of elementary excitations (and vortices) breaks the superfluid order beyond a certain velocity. Quantum friction provides a close analog to  Landau's argument in the context of macroscopic bodies. The same line of reasoning is adopted in the work of Frank and Ginzburg~\cite{Frank45,Ginzburg93,Ginzburg96}. While this argument correctly predicts the threshold velocity for the onset of friction, it does not quantify the magnitude of  friction and its dependence on system parameters.

\subsection{The input-output formalism}\label{Sec: Input-output formalism}
The input-output formalism deals with the second quantized operators corresponding to incoming and outgoing wave functions, relating them through the classical scattering matrix \cite{Matloob95,Gruner96,Loudon97,Beenakker98}. From the (known) distribution of the incoming modes, one can then determine the out-flux of the outgoing quanta.
The input-output formalism has been used to study the \emph{dynamical Casimir effect}---a consequence of the quantum field theory in the presence of moving boundaries \cite{Fulling76}---in theoretical \cite{Neto96,Lambrecht96} as well as experimental \cite{Nori11} contexts, and recently generalized in application to lossy objects by utilizing scattering techniques~\cite{Maghrebi13}. The present generalization to half-spaces in relative motion
relies on our assumption that the dielectric function is approximately a real constant.
In this context, we consider incoming waves as originated well within each half-space, far away from the gap between them (asymptotic infinity).
These waves propagate towards the gap, and then scatter (backwards or forwards) to asymptotic infinities.
The incoming and outgoing wave-functions for each half-space are normalized such that the current density perpendicular to the surface of the half-space is unity up to a sign.

The second-quantized field $\hat \Phi_i$ within the medium is indexed by $i=1,2$ designating the two objects. This operator is decomposed into modes defined by operators $a^{\out/\inn}_{i \omega \bk_\|}$ separately for incoming and outgoing waves within each half-space. Note that an
operator $a_{\omega\bk_\|}$ with negative $\omega$ should be interpreted as a
creation operator; more precisely, $a_{\omega\bk_\|}=a^\dagger_{-\omega \, -\bk_\|}$ (the momentum's sign is reversed due to  Hermitian conjugation). Crudely speaking, annihilating a negative-energy particle is equivalent to creating one with positive energy. The operator $\hat \Phi$ (with the index $i$ being implicit) is expanded as
\begin{align}
    \hat \Phi= \sqrt{\frac{\hbar}{2}}\sum_{\omega \bk_\|} e^{-i\omega t}(\varphi^{\inn}_{\omega \bk_\|} \hat a^{\inn}_{\omega \bk_\|} + \varphi^{\out}_{\omega \bk_\|} \hat a^{\out}_{\omega \bk_\|})+\mbox{h.c.}\, ,
\end{align}
with two copies of this equation, one for each half-space. The wavefunction $\varphi_{\omega \bk_\|}$, in the object's rest frame, is defined as
  \begin{align}\label{Eq: In and Out Fns}
    \varphi_{\omega \bk_\|}^{\out /\inn }= \Big\{ \begin{array}{cc}
      \frac{1}{\sqrt{\tilde k_\perp}} e^{i \bk_\|\cdot \bx_\| \pm i \tilde k_\perp z}, \hskip .3in  \omega>0,\\
      \frac{1}{\sqrt{\tilde k_\perp}} e^{i \bk_\|\cdot \bx_\| \mp i \tilde k_\perp z}, \hskip .3in  \omega<0,
      \end{array}
  \end{align}
with the prefactor being chosen to ensure the normalization, $z$ measuring the distance from the surface, and $\tilde k_\perp=\sqrt{\epsilon \, \omega^2/c^2-\bk_\|^2}$. Note that the designation of incoming or outgoing for
propagating waves depends on the relative signs of $\omega$ and $\tilde k_\perp$, as indicated in the above equation. For the moving half-space, the corresponding wavefunctions are obtained by a Lorentz transformation of $\omega$ and $k_x$, while $\tilde k_\perp$ (being perpendicular to the velocity) remains invariant.

With the operators defined above, the input-output relation takes the form
  \begin{align}\label{Eq: S matrix}
    \left( \begin{array}{cc}
    \hat a^{\out}_1 \\
    \hat a^{\out}_2   \end{array} \right)= \mathbb S \,
    \left( \begin{array}{cc}
    \hat a^{\inn}_1   \\
    \hat a^{\inn}_2   \end{array} \right),
  \end{align}
where $\mathbb S$ is the $2\times 2$ scattering matrix, and the dependence on $\omega$ and $\bk_\|$ is implicit. The scattering matrix can be straightforwardly computed by matching the wavefunction and its first derivatives along the boundaries. Note that a scattering channel relates a wavefunction labeled by $(\omega, \bk_\|)$ on one half-space to $(\omega',\bk_\|')$, the frequency and wavevector as seen from the moving frame, on the second half-space. At small velocities, we have $\omega'\approx\omega-v k_x$ and $\bk_\|'\approx\bk_\|$. Therefore, a positive-frequency mode on one half-space can be coupled to a mode with negative frequency on the other half-space. However, as remarked above, an operator $a_{\omega\bk_\|}$ with negative $\omega$ is, in fact, a creation operator. This mixing between positive and negative frequencies is at the heart of the dynamical Casimir effect~\cite{Dalvit11,Maghrebi13}. In a frequency window where this mixing occurs, i.e. for $0<\omega<v k_x$, the input-output relation is recast as
  \begin{equation}\label{Eq: mixing}
    \hat a_{1\,\omega \bk_\|}^{\out}=S_{11} \, \hat a_{1\, \omega \bk_\|}^{\inn}+S_{21} \, \hat a_{2\, vk_x\!\!-\omega \,-\bk_\|}^{\inn\, \dagger}\,.
  \end{equation}
  One can then compute the expected flux of the outgoing modes, $\langle \hat a^{\out \,\dagger}_1 \hat a^{\out }_1\rangle$. At zero temperature, $\langle \hat a^{\inn \,\dagger}_1 \hat a^{\inn }_1\rangle=0$, and the only contribution to the outflux is due to the second term in the RHS of Eq.~(\ref{Eq: mixing}), resulting in
  \begin{align}
    \langle \hat a^{\out \,\dagger}_{1\, \omega \bk_\|} \hat a^{\out }_{1\, \omega \bk_\|}\rangle= \Theta(v k_x -\omega)\,\, |S_{21}|^2,
  \end{align}
  with $\Theta$ being the Heaviside step function.
The friction, or the rate of the lateral momentum transfer from one half-space to another, is then
  \begin{align}\label{Eq: L mom Rad}
    f=&\int_{0}^{\infty} {\dbar}\omega \, L^{D-1}\!\int{\dbar} \bk_\| \,\, \hbar k_x \, \Theta(v k_x -\omega)\,\, |S_{21}|^2\,.
  \end{align}
Similarly, the energy radiation can be computed by replacing $\hbar k_x$ by $\hbar \omega$ in the last equation.

It is a simple exercise to compute the scattering matrix: Exploiting the continuity equations (of the field and its first derivative) at the interface of the two half-spaces and the gap, we find the set of equations
\begin{align}\label{Eq: S-{21}}
 & 1+S_{11}=A+B,  && i \tilde k_\perp (1-S_{11})  = |k_\perp|\,(A-B),
 \nonumber \\
 & A e^{|k_\perp| d}+B e^{-|k_\perp| d}=\sqrt{\frac{\tilde k_\perp}{\tilde k_\perp'}}\,S_{21} e^{i \tilde k'_\perp}, && |k_\perp|\, (A e^{|k_\perp| d}-B e^{-|k_\perp| d})=i \sqrt{{\tilde k_\perp}{\tilde k_\perp'}}\,S_{21} e^{i \tilde k_\perp' d},
\end{align}
where $k_\perp$ and $\tilde k_\perp$ are defined above, $\tilde k'_\perp=\sqrt{\epsilon \, \omega'^2-\bk_\|'^2}$, and $A$ and $B$ are
appropriate coefficients to be determined. Notice that the waves in the medium are propagating while those within the gap are evanescent, consistent with the constraints outlined above. One can then solve for $S_{21}$ explicitly, and, using Eqs.~(\ref{Eq: reflection matrix}) and (\ref{Eq: S-{21}}), show that (for real $\epsilon$)
\begin{equation}
  |S_{21}|^2=    \frac{e^{-2 |k_\perp| d}  \, (2 \im R_1) \,(2\im R_2)}{{|1-e^{-2|k_\perp| d} R_1 R_2|^2}}.
\end{equation}
(Note that $R_1=- (\tilde k_\perp -i|k_\perp|)/(\tilde k_\perp +i|k_\perp|)$ and similarly for $R_2$ with $\tilde k_\perp\to \tilde k_\perp'$.)
It is immediately clear that Eq.~(\ref{Eq: L mom Rad}) reproduces the friction in Eq.~(\ref{Eq: Friction}). Indeed the input-output formalism makes the derivation rather trivial.
\begin{figure}[h]
    \centering
 \includegraphics[width=14cm]{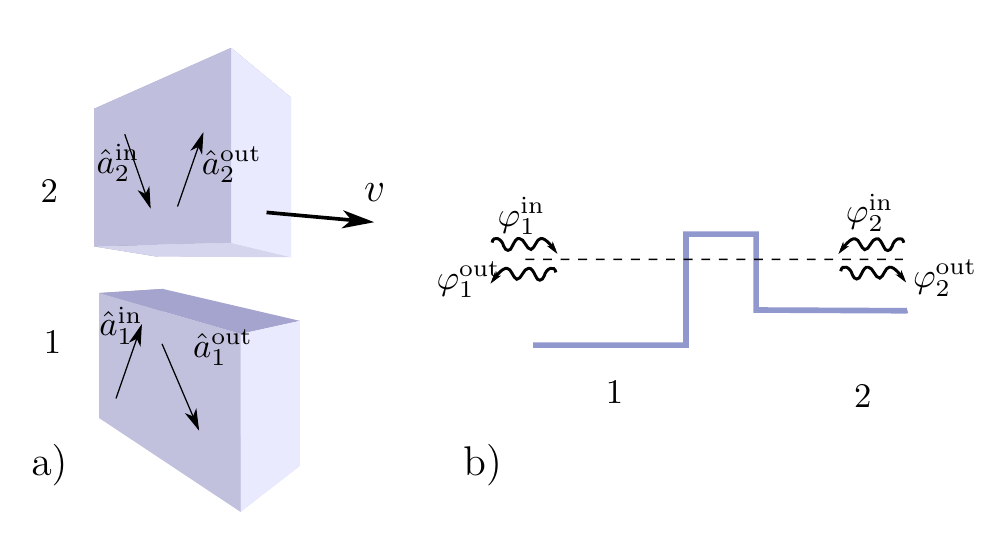}
   \caption{a) The operators within each object represent incoming and outgoing modes, related in the input-output formalism through the scattering matrix. b) The scattering problem is similar to the quantum tunneling over a barrier; friction resulting from transfer of momentum by the tunneling quanta.} \label{Fig: Input-Output}
 \end{figure}

Equations (\ref{Eq: S matrix}) and (\ref{Eq: mixing}) relate annihilation and creation operators which satisfy canonical commutation relations,
  \begin{align}
    [\hat a^{\out/\inn }_{i \, \omega \bk_\|},  \hat a^{\out/\inn \,\dagger}_{j \, \omega \bk_\|}]=\sgn(\omega)\,\delta_{ij}\,.
  \end{align}
The function $\sgn(\omega)$ merely indicates that, for negative frequencies, the creation and annihilation operators should be identified correctly.
The above canonical relations applied to Eq.~(\ref{Eq: mixing}) yield
  \begin{equation}\label{Eq: super unitary}
    1-|S_{11}|^2 =\sgn(\omega - v k_x) |S_{21}|^2  \,,
  \end{equation}
implying that the scattering amplitude corresponding to the backscattering in the first medium is larger than unity for $\omega< v k_x$. This is an example of the so-called \emph{superscattering }due to Zel'dovich~\cite{Zel'dovich71}: For certain modes, a moving (rotating) object amplifies incoming waves indicating that energy is extracted from motion. A closely related phenomenon occurs for rotating black holes and is known as the Penrose process~\cite{Penrose69}. Superradiance signals quantum instability of the moving object resulting in energy and momentum radiation, and a corresponding exertion of frictional force~\cite{Zel'dovich71,Bekenstein98,Maghrebi12}.\footnote{An alternative approach to the rotational friction is given in Ref.~\cite{Manjavacas10}.} Equations~(\ref{Eq: L mom Rad}) and (\ref{Eq: super unitary}) can be combined to yield
\begin{equation}
   {\cal P}=\int_{0}^{\infty} {\dbar}\omega \, L^{D-1}\!\int{\dbar} \bk_\| \,\, \hbar \omega  \, \Theta(v k_x -\omega)\,\, (|S_{11}|^2-1)\,.
\end{equation}
This expression is indeed very similar to quantum radiation from a rotating object (a rotating black hole in Ref.~\cite{Unruh74} or a rotating cylinder in Ref.~\cite{Maghrebi12}) with the following substitutions: $v\to \Omega$ (linear to angular velocity), $k_x \to m$ (linear to angular momentum), and $S_{11}$ to be replaced by the scattering matrix of the rotating body.

It is worth noting that the only contribution to the radiation is from modes with $k_x> \omega/v> \omega/c$, corresponding to evanescent waves in the gap between the two half-spaces. In this respect, the radiation is a quantum  tunneling process across a barrier (in this case, the gap). We can recast the wave equation in a fashion similar to the Schr\"{o}inger equation as
  \begin{equation}
    (-\partial_z^2+ V)\varphi=0\,,
  \end{equation}
  with $V_1=-(\epsilon \, {\omega^2}/{c^2}-\bk_\|^2)$, $V_2=-(\epsilon \, {(\omega-v k_x)^2}/{c^2}-\bk_\|^2)$, and $V_{\rm gap}=|{\omega^2}/{c^2}-\bk_\|^2|$; see Fig.~\ref{Fig: Input-Output}.
  The relative motion of the two media results in a steady tunneling of particles of opposite momenta from one half-space to another, thus leading to the slowdown of the motion.

\subsection{Inner-product method} \label{Sec: Inner-product method}
  In this section, we describe a method which is widely used in application to quantum field theory in curved space-time, or in the presence of moving bodies. However, the following discussion is, to the best of our knowledge, the first application of this method to the problem of moving half-spaces,
which is possible only when the dielectric function is taken to be a real constant.

  To quantize a field theory, a first step is to decompose the quadratic part of the Hamiltonian into a collection of (infinite) harmonic oscillators; define the corresponding annihilation and creation operators, and impose canonical commutation relations. One can then construct the Fock space with the vacuum state of no particles, and excited single- and multi-particle states obtained by applying creation operators. In high energy physics, the usual starting point is empty space, but the above procedure works equally well in the presence of background matter, as is the case for the Casimir effect. The reason is that canonical quantization only relies upon time translation and time reversal symmetry. The former allows construction of eigenmodes of a definite frequency, which is the basis of the notion of modes/quanta/particles. Time reversal symmetry, on the other hand, is used to identify creation and annihilation operators: The coefficient of a positive (negative) frequency mode is understood as an annihilation (creation) operator. To make this correspondence explicit, consider a quantum field $\hat \Phi (t,\bx)$, possibly in the presence of a background \emph{medium} which is static. One can find a basis of eigenmodes $\phi_{\omega \alpha}(\bx)$ labeled by frequency $\omega$ and quantum number $\alpha$ to expand the field as
  \begin{equation}\label{Eq: Expansion}
    \hat \Phi(t,\bx)=\sqrt{\frac{\hbar}{2}}\sum_{\omega>0, \, \alpha} e^{-i\omega t}\phi_{\omega \alpha}(\bx) \, \hat a_{\omega \alpha} + e^{i\omega t}\phi^*_{\omega \alpha}(\bx) \, \hat a^\dagger_{\omega \alpha}\,,
  \end{equation}
  with (defining $\ddelta(x)=2\pi\delta(x)$)
  \begin{equation}
    [a_{\omega_1 \alpha}, a^\dagger_{\omega_2\beta}]=\ddelta(\omega_1-\omega_2)\, \delta_{\alpha\beta}\,.
  \end{equation}
  The latter follows from the canonical commutation relations between the field $\hat \Phi(t,\bx)$ and its conjugate momentum $\hat \Pi(t,\by)$,
  \begin{equation}
    [\hat\Phi(t,\bx), \hat\Pi(t,\by)]=i \hbar \, \delta(\bx-\by).
  \end{equation}

When the object is moving, we lose one or both symmetries in time. The case of two parallel plates in lateral motion respects time translation symmetry as the relative position does not change. Time reversal symmetry, on the other hand, is broken; in the backward direction of time the half-space moves in the opposite direction. In the absence of time reversal symmetry, the correspondence between positive (negative) frequency and the annihilation (creation) operators breaks down. There is, however, a more general way to identify operators as follows. Let us consider two functions $\phi_1$ and $\phi_2$, which are solutions to the classical field equation, and define an inner product as~\cite{birrell1984, fulling1989, Carroll04}
  \begin{equation}\label{Eq: inner product}
    \langle \phi_1, \phi_2 \rangle=\frac{i}{2}\int \dd\bx\,(\phi_1^* \pi_2-\pi_1^* \phi_2),
  \end{equation}
where $\pi_i$ is the corresponding conjugate momentum, and the integral is over the whole space. One can easily see that the inner product defined in Eq.~(\ref{Eq: inner product}) is independent of the choice of the reference frame or the (space-like) hypersurface as the integration domain. Furthermore, we can always find a set of functions $\phi_{\omega \alpha}$ which solve the classical field equation and form an  orthonormal basis,
  \begin{equation}
    \langle \phi_{\omega_1 \alpha}, \phi_{\omega_2 \beta}\rangle =\ddelta(\omega_1-\omega_2)\delta_{\alpha\beta},
  \end{equation}
  while their conjugate modes are orthonormal up to a negative sign,
  \begin{equation}\label{Eq: norm 1}
    \langle \phi^*_{\omega_1 \alpha}, \phi^*_{\omega_2 \beta}\rangle =-\ddelta(\omega_1-\omega_2)\delta_{\alpha\beta}.
  \end{equation}
Note that $\omega$ is still a good quantum number because of the translation symmetry in time.
These modes form a complete basis, such that the field $\hat\Phi$ can be expanded as
  \begin{equation}\label{Eq: norm 2}
    \hat \Phi(t,\bx)=\sqrt{\frac{\hbar}{2}}\sum_{\omega\alpha:\,\mbox{\scriptsize positive-norm}} e^{-i\omega t}\phi_{\omega \alpha}(\bx) \, \hat a_{\omega \alpha} + e^{i\omega t}\phi^*_{\omega \alpha}(\bx) \, \hat a^\dagger_{\omega \alpha}\,.
  \end{equation}
  Therefore, an annihilation (creation) mode should be more generally identified with a positive (negative) norm and not frequency; the latter depends on the reference frame while the former does not. The above relations become obvious for a field theory in a static background where the conjugate momentum is proportional to $\partial_t \Phi$; Eq.~(\ref{Eq: inner product}) then implies that a positive (negative) $\omega$ corresponds to positive (negative) norm.

  The inner-product method is used in application to quantum field theory in curved space-time. For example, it has been employed to show that a rotating black hole is unstable due to spontaneous emission~\cite{Unruh74}.
For two parallel plates in motion, we first introduce a complete basis. Due to the translation symmetry in time and space (parallel to the half-spaces' surface), wavefunctions can be labelled by frequency $\omega$ and tangential wavevector $\bk_\|$ defined in the lab frame where the first half-space is at rest and the second one is moving. There are two independent solutions defined as
  \begin{align}
      \phi^{\rm I}_{\omega\bk_\|}= {\Big\{}\begin{array}{cc}
     \varphi^{\inn}_{1 \, \omega\bk_\|}+S_{11}\,\varphi^{\out}_{1\, \omega\bk_\|}, \hskip .2in \mbox{half-space 1},\\
      S_{21}\,\varphi^{\out}_{2\, \omega\bk_\|}, \hskip .8in \mbox{half-space 2},
      \end{array}
  \end{align}
  and
  \begin{align}
    \phi^{\rm II}_{\omega\bk_\|}= {\Big\{}\begin{array}{cc}
     S_{12}\,\varphi^{\out}_{1\, \omega\bk_\|}, \hskip .8in \mbox{half-space 1},\\
      \varphi^{\inn}_{2\, \omega\bk_\|}+S_{22}\,\varphi^{\out}_{2\, \omega\bk_\|} ,\hskip .2in \mbox{half-space 2}.\end{array}
  \end{align}
  The incoming and outgoing functions are defined in Eq.~(\ref{Eq: In and Out Fns}), with the functions in the second half-space properly Lorentz-transformed; see the explanation below Eq.~(\ref{Eq: In and Out Fns}). To find the conjugate momentum, note that Eq.~(\ref{Eq. field equation}) can be schematically derived from a Lagrangian ${\cal L}=\frac{1}{2}[\frac{1 }{v_0^2}(\partial_t \Phi)^2-(\nabla \Phi)^2]$ with $v_0=c/\sqrt{\epsilon}$, hence $\Pi={\partial L}/{\partial \dot\Phi}=\frac{1}{v_0^2} \,\partial_t\Phi$. Similarly for a moving object ${\cal L}\approx \frac{1}{2}[\frac{1 }{v_0^2}(\partial_t \Phi+v \partial_x \Phi)^2-(\nabla \Phi)^2]$ and $\Pi=\frac{1}{v_0^2}(\partial_t \Phi+v \partial_x \Phi)$. In terms of partial waves, $\pi_{\omega \bk}= -i\frac{\omega-vk_x}{v_0^2}\,\phi_{\omega\bk}$ within the moving half-space and similarly for the static half-space with $v=0$.
  One can then see that
  \begin{equation}\label{Eq: normlz 1}
    \langle \phi^{\rm I}_{\omega_1\bk_\|}, \phi^{\rm I}_{\omega_2\bl_\|}\rangle =\ddelta(\omega_1-\omega_2)\, \ddelta(\bk_\|-\bl_\|), \hskip 1.1in \omega_1>0,
  \end{equation}
  and
  \begin{equation}\label{Eq: normlz 2}
    \langle \phi^{\rm II}_{\omega_1\bk_\|}, \phi^{\rm II}_{\omega_2\bl_\|}\rangle =\sgn (\omega_1-v k_x)\, \ddelta(\omega_1-\omega_2)\, \ddelta(\bk_\|-\bl_\|), \hskip .2in \omega_1>0.
  \end{equation}
  To obtain these relations, we have exploited the fact that the norms are diagonal in frequency to compute the delta functions in $\tilde k_\perp$ from Eqs.~(\ref{Eq: normlz 1}) and (\ref{Eq: normlz 2}) which are then converted to those of $\omega$. The integral over the gap is neglected as it does not contribute to the frequency delta functions. Note that the (super)unitary relation in Eq.~(\ref{Eq: super unitary}) is essential in deriving the norms. Functions of type I have positive norm so they serve as the coefficients of annihilation operators. However, type II functions include negative-norm modes for $0<\omega< v k_x$. Therefore, despite the positive sign of frequency, the latter should be identified as creation operators. We thus expand the field as
\begin{align}
     \frac{\hat \Phi(t,\bx)}{\sqrt{{\hbar}/{2}}}=&\sum_{\omega>0, \, \bk_\|} e^{-i\omega t} \phi^{\rm      I}_{\omega \bk_\|}(\bx) \, \hat a^{\rm I}_{\omega \bk_\|}+\cr
     &\sum_{0<\omega\,, v k_x < \omega \,, \bk_\|} e^{-i\omega t}  \phi^{\rm II}_{\omega \bk_\|}(\bx) \, \hat a^{\rm II}_{\omega \bk_\|} +\sum_{ 0<\omega<v k_x, \, \bk_\|} e^{-i\omega t} \phi^{\rm II}_{\omega \bk_\|}(\bx) \, \hat a^{{\rm II} \, \,\dagger}_{\omega \bk_\|}\quad +\,\,{\rm h.c.}\,,
\end{align}
where the summation is a shorthand for multiple integrals, and $\hat a$ and $\hat a^\dagger$ satisfy the usual commutation relations.
   The friction is given by the rate of the lateral momentum transfer,
   \begin{align}\label{Eq: radiation from norm}
   \frac{f}{L^{D-1}} &  = \langle \partial_x \Phi \, \partial_z \Phi \rangle \cr
     &=\int_{0}^{\infty} {\dbar} \omega \!\int{\dbar} \bk_\| \,\frac{\hbar \,k_x }{2}\Big\{ -(1-|S_{11}|^2)+ |S_{12}|^2 \Big\}\cr
     &=\int_{0}^{\infty} {\dbar} \omega \!\int{\dbar} \bk_\| \,\, \hbar k_x\, \Theta(v k_x -\omega)\,\, |S_{12}|^2\,.
   \end{align}
We have again exploited Eq.~(\ref{Eq: super unitary}), and arrived at the same results as in the previous sections. Notice that only the \emph{superradiating modes} contribute to the radiation while other modes cancel out in the second line of the last equation.

\subsection{Radiated energy: The Rytov formalism}\label{Sec: Friction vs radiation}
In this section, we employ the \emph{Rytov }formalism~\cite{Rytov89} to study the correspondence between friction and radiation in some detail. This formalism is based on the fluctuation-dissipation theorem, and has been extensively used in the context of non-contact friction~\cite{Volokitin99}; see also Ref.~\cite{Volokitin07} and citations therein. This section goes beyond the existing literature by computing various correlation functions, and the radiated energy  \emph{inside} the half-spaces (as well as in the gap between them), thus making an explicit connection to  Cherenkov radiation. Specifically, we discuss the dependence of various quantities on the reference frame: While the radiation in the gap depends on the reference frame (in the center of mass frame, the latter is simply zero), the radiation within the two half-spaces is invariant and presents a close analog to classical Cherenkov radiation.

We start by relating  fluctuations of the field to those of ``sources'' within each
medium by
  \begin{align}\label{Eq: Rytov for scalar field}
  -\left(\triangle + \frac{\omega^2}{c^2} \, \epsilon(\omega, \bx)\right)\Phi(\omega, \bx)=- \frac{i \omega}{c} \, \rho_\omega (\bx)\,.
  \end{align}
The  ``charge" $\rho$ fluctuates around zero mean with correlations (co-variance)
\begin{equation}
 \langle \rho_\omega(\bx) \rho^*_\omega(\by) \rangle =a(\omega)\im \epsilon(\omega, \bx) \, \delta(\bx-\by),
  \end{equation}
where
  \begin{equation}
  a(\omega)= 2\hbar \left[n(\omega,T)+\frac{1}{2}\right]= \hbar \, \coth\left(\frac{\hbar \omega}{2k_B T}\right)\,.
  \end{equation}
Note that the source term on the RHS of Eq.~(\ref{Eq: Rytov for scalar field}) comes with a coefficient linear in frequency reminiscent of a time derivative, the reason being that the source couples to the time derivative of the field just in the same way that the response function, $\epsilon$, correlates the time derivatives of the field at different times.

The field is related to the sources via the Green's function, $G$, defined by
  \begin{equation}\label{Eq: Green's function}
    -\left(\triangle + \frac{\omega^2}{c^2} \epsilon(\omega, \bx)\right)G(\omega, \bx,\bz)=\delta(\bx-\bz).
  \end{equation}
In equilibrium (uniform temperature and static), this results in the field correlations
  \begin{align}\label{Eq: eq FDT}
    \langle \Phi(\omega, \bx)\Phi^*(\omega, \by)\rangle
    &= \frac{\omega^2}{c^2} \int_{\mbox{\scriptsize All space}}
  \dd\bz  \,G(\omega, \bx,\bz) G^*(\omega, \by,\bz) \, \langle \rho_\omega(\bz) \rho^*_\omega(\bz) \rangle \nonumber\\
    &= \frac{\omega^2}{c^2} a(\omega) \int_{\mbox{\scriptsize All space}} \hskip -.3in \dd \bz \, G(\omega,\bx,\bz) \im \epsilon (\omega,\bz) \, G^*(\omega,\by,\bz) \nonumber \\
    &= a(\omega) \im G(\omega, \bx,\by),
  \end{align}
in agreement with the fluctuation-dissipation condition which relates correlation functions to dissipation through the imaginary part of the response function. Note that the second line in Eq.~(\ref{Eq: eq FDT}) follows from $\frac{\omega^2}{c^2}\im\epsilon =-{\im G^{-1}}$ according to Eq.~(\ref{Eq: Green's function}).

For the case of two half-spaces, we first compute the correlation function for two points in the gap. The source fluctuations in each half-space will be treated separately;
starting with those in the static half-space (indicated by  sub-index 1 on the integral)
  \begin{align}\label{Eq: GF--Fluc in 1}
   \langle \Phi(\omega, \bx)&\Phi^*(\omega, \by)\rangle_1
  =\frac{\omega^2}{c^2} a_1(\omega) \int_{1}
  \dd\bz  \, G(\omega, \bx,\bz) \im \epsilon(\omega, \bz) \, G^*(\omega, \by,\bz)\nonumber\\
  &=\frac{\omega^2}{2 i c^2} a_1(\omega) \int_{1}
  \dd\bz  \, \left[\epsilon(\omega, \bz) G(\omega, \bx,\bz) \right]\, G^*(\omega, \by,\bz)-G(\omega, \bx,\bz) \, \left[\epsilon(\omega, \bz) G(\omega, \by,\bz)\right]^*\nonumber\\
  &=\frac{i}{2} a_1(\omega) \int_{1}
  \dd\bz  \, \left[\triangle_{\bz} G(\omega, \bx,\bz) \right]\, G^*(\omega, \by,\bz)-G(\omega, \bx,\bz) \, \triangle_{\bz} G^*(\omega, \by,\bz)\nonumber\\
  &=\frac{i}{2} a_1(\omega) \int_{1} \dS\cdot \left[ (\nabla_{\bz} G(\omega, \bx,\bz) )\, G^*(\omega, \by,\bz)-G(\omega, \bx,\bz) \, \nabla_{\bz} G^*(\omega, \by,\bz)\right]\,.
  \end{align}
  Note that we used Eq.~(\ref{Eq: Green's function}) in going from the second to the third line above, and then integrated by parts to obtain an integral over the surface adjacent to the gap. The contribution due to the other surface at infinity vanishes since $\epsilon$ is assumed to have a vanishingly small imaginary part. This assumption is a rather technical point which also arises for the dielectric response of the vacuum in the context of a single object out of thermal or dynamical equilibrium with the vacuum \cite{Kruger11, Maghrebi12}.

To compute the surface integral in Eq.~(\ref{Eq: GF--Fluc in 1}), one needs the (out-out) Green's function with both points in the gap.
The latter is given by Eq.~(\ref{Eq: G out-out}), and leads to
  \begin{align}
    \langle \Phi(\omega, \bx)\Phi^*(\omega, \by)\rangle_1& =
    - \sum_\alpha \frac{a_1(\omega) }{4 p_\alpha^*} \frac{|e^{i p_\alpha d}|^2}{|1- e^{2i p_\alpha d}R_\alpha \tilde R_\alpha|^2}  \, |U(R_\alpha)|^2 \times \nonumber \\     &
    \left(\Phi^{\reg}_\bbalpha(\omega, \tilde\bx) +\tilde R_\bbalpha \Phi_\bbalpha^{\out}(\omega,\tilde \bx)\right) \overline{    \left(\Phi^{\reg}_\bbalpha(\omega, \tilde\by) +\tilde R_\bbalpha \Phi_\bbalpha^{\out}(\omega,\tilde \by)\right)}\,.
  \end{align}
In this equation, $R_\alpha$ ($\tilde R_\alpha$) is the reflection coefficient from the first (second) object with $\alpha$ being a shorthand for both the frequency and  wavevector. Also $\bx/\by$ ($\tilde \bx/\tilde \by$) is the distance from a reference point on the surface of the first (second) half-space---the reference points on two surfaces have identical parallel components, $\bx_\|=\tilde\bx_\|$, but differ in their $z$-component as $z_{\tilde\bx}=d-z_\bx$.
The regular and outgoing functions are defined with respect to the corresponding half-space; see Appendix~A for more details. Furthermore, the overbar notation implies complex conjugation, and $|U(R_\alpha)|^2$ is defined as
  \begin{equation}\label{Eq: normalization1}
    |U(R_\alpha)|^2=\frac{\int \dS  \cdot \left[{\Phi^*_\alpha} \, \nabla {\Phi_\alpha}-(\nabla {\Phi^*_\alpha}) \, {\Phi_\alpha}\right]}{\int \dS  \cdot \left[{\Phi_\alpha^{\reg *}} \, \nabla {\Phi^{\reg}_\alpha}-(\nabla {\Phi^{\reg *}_\alpha}) \, {\Phi^{\reg}_\alpha}\right]}\,,
  \end{equation}
  with $\Phi_\alpha=\Phi_\alpha^{\reg}(\omega,\bz)+R_\alpha \Phi_{\alpha}^{\out}(\omega,\bz)$, such that
  \begin{align}
|U(R_\alpha)|^2
      &=1 -|R_\alpha|^2, \qquad \alpha \in{\mbox{propagating waves}},\cr
      &=2 \im R_\alpha, \qquad\quad  \alpha\in{\mbox{evanescent waves}}.
  \end{align}
  One can similarly find the correlation function due to source fluctuations in the second half-space
  \begin{align}
     \langle \Phi(\omega, \bx)\Phi^*(\omega, \by)\rangle_2&=
    -\sum_\alpha \frac{a_1(\omega-\bv \cdot \bk_\alpha) }{4 p_\alpha^*} \frac{|e^{i p_\alpha d}|^2}{|1- e^{2i p_\alpha d}R_\alpha \tilde R_\alpha|^2} \, |U(\tilde R_\alpha)|^2 \times \nonumber \\     &
    \left(\Phi^{\reg}_\bbalpha(\omega, \bx) +R_\bbalpha \Phi_\bbalpha^{\out}(\omega,\bx)\right) \overline{    \left(\Phi^{\reg}_\bbalpha(\omega, \by) +R_\bbalpha \Phi_\bbalpha^{\out}(\omega,\by)\right)}\,\,.
  \end{align}
  The total correlation function is the sum of the contributions due to each half-space,
  \begin{align}
      \langle \Phi(\omega, \bx)\Phi^*(\omega, \by)\rangle
      =  \langle \Phi(\omega, \bx)\Phi^*(\omega, \by)\rangle_1+  \langle \Phi(\omega, \bx)\Phi^*(\omega, \by)\rangle_2. \nonumber
  \end{align}
The frictional force is then computed as the average of the appropriate component of the stress tensor as
  \begin{align} \label{Eq: friction 2}
     & f   =\int_{-\infty}^{\infty} {\dbar} \omega\int \dd\Sigma\, \langle \partial_x \Phi(\omega, \bx) \partial_z \Phi^*(\omega, \bx)\rangle \nonumber \\
     &  =\int_{0}^{\infty} \!\!{\dbar} \omega\,L^{D-1}\!\!\!\int {{\dbar}\bk_\|} \,\hbar k_x \,\frac{|e^{i k_\perp d}|^2}{{|1- e^{2i k_\perp d}R_{\omega\bk_\|}\tilde R_{\omega\bk_\|}|^2}} |U(R_{\omega\bk_\|})|^2 |U(\tilde R_{\omega\bk_\|})|^2 \, (n_1(\omega)-n_2(\omega-\bv\cdot \bk)),
  \end{align}
where we have restored $\bk_\|$ in place of $\alpha$.
Further manipulations lead to Eq.~(\ref{Eq: Friction}).
Similarly, the Rytov  formalism allows us to calculate the energy flux from one object to the other as
  \begin{align}\label{Eq: P-gap}
   &  P_{\rm gap}   =\int_{-\infty}^{\infty} {\dbar} \omega\int \dd\Sigma\, \langle \partial_t \Phi(\omega, \bx) \partial_z \Phi^*(\omega, \bx)\rangle \nonumber \\
   &   =\int_{0}^{\infty} \!\!\! {\dbar} \omega\,L^{D-1}\!\!\!\int {{\dbar}\bk_\|} \,\,\hbar \omega \,\frac{|e^{i k_\perp d}|^2}{{|1- e^{2i k_\perp d}R_{\omega\bk_\|}\tilde R_{\omega\bk_\|}|^2}} |U(R_{\omega\bk_\|})|^2 |U(\tilde R_{\omega\bk_\|})|^2 \, (n_1(\omega)-n_2(\omega-\bv \cdot \bk)),
  \end{align}
{i.e.} by merely  replacing $\hbar k_x$ with $\hbar \omega$ in Eq.~(\ref{Eq: friction 2}).

Next we compute the energy flux {\it through each half-space}.
Since the dielectric function is assumed to be a real constant (albeit with a vanishingly small imaginary part), we can circumvent ambiguities in defining the Maxwell stress tensor in a lossy medium \cite{Jackson98}.
In the following, we find the field correlation function in the first (static) half-space due to source fluctuations in the moving half-space, by using an analog of Eq.~(\ref{Eq: GF--Fluc in 1})  but evaluating a surface integral on the second half-space. However,  in this case, the appropriate Green's function is the (out-in) type given in Eq.~(\ref{Eq: G out-in}). We then find the latter correlation function as
   \begin{align}
     \langle \Phi(\omega, \bx)\Phi^*(\omega, \by)\rangle_2&=
    \sum_\alpha \frac{a_2(\omega-\bv \cdot \bk_\alpha)}{4 p_\alpha^*} \frac{|e^{i p_\alpha d}|^2}{|1- e^{2i p_\alpha d}R_\alpha \tilde R_\alpha|^2} \, |U(\tilde R_\alpha)|^2 \times \nonumber \\
    &\left(V_{\alpha}\psi_\alpha^{+}(\omega,\bx)+W_\alpha \psi_{\alpha}^{-}(\omega,\bx)\right)
    \overline{\left(V_{\alpha}\psi_\alpha^{+}(\omega,\by)+W_\alpha \psi_{\alpha}^{-}(\omega,\by)\right)}\,\,,
  \end{align}
  where $\bx$ and $\by$ are both inside the first half-space, $V$ and $W$ are coefficients depending on $\alpha$ and system parameters, and the functions $\psi$ are defined inside the medium; see Appendix~A for more details.
  Henceforth, we assume that the objects are at zero temperature. Anticipating that only evanescent waves contribute, we obtain the energy flux in the first half-space due to the fluctuations in the second half-space. Noting that the ``Poynting vector'' is defined as $\partial_t \Phi \partial_z\Phi$ even within the dielectric medium, we find
  \begin{align}
     P^{(1)}_2=\int{\dbar} \omega\,L^{D-1}\!\!\!\int {{\dbar}\bk_\|} \,\,\hbar \omega \,\frac{e^{-2 |k_\perp| d}}{{|1- e^{-2 |k_\perp| d}R_{\omega\bk_\|} \tilde R_{\omega\bk_\|}|^2}} \, 2 \im R_{\omega\bk_\|} \, 2 \im \tilde R_{\omega\bk_\|} \, \sgn(\omega -\bv \cdot \bk),
  \end{align}
where we have used the fact that, for evanescent waves, $p_\alpha\equiv k_\perp=\sqrt{\omega^2/c^2-\bk_\|^2}$ is purely imaginary while $\tilde p_\alpha\equiv \tilde k_\perp= \sqrt{\epsilon \, \omega^2/c^2-\bk_\|^2}$ is real, leading to
  \begin{align}
    |V_\alpha|^2-|W_\alpha|^2=\frac{|p_\alpha|}{\tilde p_\alpha} \, 2 \im R_{\alpha}\,.
  \end{align}
  In order to take into account the source fluctuations in the first half-space (where we compute the field correlation function), we need the (in-in) Green's function in Eq.~(\ref{Eq: G in-in}).
  The energy flux due to the latter fluctuations, $P^{(1)}_1$, is computed similarly but there is one subtlety. Unlike the previous cases, the correlation function is evaluated at points where there are also fluctuating sources. However, Eq.~(\ref{Eq: GF--Fluc in 1}) contains, beyond the surface integral, a term proportional to $\im G(\omega,\bx,\by)$ which does not contribute to the radiation. The remaining computation is similar to the previous case, and the overall energy flux is obtained as
  \begin{align} \label{Eq: P (1)}
     P^{(1)} &  =P^{(1)}_1+P^{(1)}_2 \nonumber\\
     &=\!\!\!\int_{0}^{\infty}{\dbar} \omega\,L^{D-1}\!\!\!\int {{\dbar}\bk_\|} \,\,\hbar \omega \,\frac{e^{-2 |k_\perp| d}\, 2 \im R_{\omega\bk_\|} \, 2 \im \tilde R_{\omega\bk_\|}}{{|1- e^{2i k_\perp d}R_{\omega\bk_\|} \tilde R_{\omega\bk_\|}|^2}}  \, \Theta(\bv \cdot \bk -\omega).
  \end{align}
This is again in harmony with the results in the previous sections.

Comparing Eqs.~(\ref{Eq: P (1)}) and~(\ref{Eq: P-gap}), we observe that in the reference frame in which the first half-space is at rest,
  \begin{equation}
    P^{(1)}=P_{\rm gap}.
  \end{equation}
  However, $P_{\rm gap}$ must vanish in the \emph{center of mass} (c.m.) frame from symmetry considerations. It can be obtained explicitly by a Lorentz transformation from the lab frame, which, to the lowest order in velocity, takes the form
  \begin{equation}
    0=P^{\rm c.m.}_{\rm gap}= P_{\rm gap}- v f/2,
  \end{equation}
indicating $P^{(1)}=vf/2$. This conclusion can be verified directly as follows: First note that Eqs.~(\ref{Eq: friction 2}) and (\ref{Eq: P (1)}) yield
  \begin{align}
    P^{(1)}-\frac{vf}{2}=\int_{k_x>0} {{\dbar}\bk_\|} \int_{0}^{v k_x}{\dbar} \omega\,\hbar \left(\omega-\frac{v k_x}{2}\right) \,\frac{e^{-2 |k_\perp| d}\, 2 \im R_{\omega\bk_\|} \, 2 \im \tilde R_{\omega\bk_\|}}{{|1- e^{2i k_\perp d}R_{\omega\bk_\|} \tilde R_{\omega\bk_\|}|^2}} ,
  \end{align}
where the $x$ axis is chosen parallel to the velocity $\bv$.
Let us make the following change of variables
  \begin{align} \omega'&=\omega-v k_x/2, \cr
      k_x' &=k_x- v \omega/2c^2\approx k_x, \cr
      k_i'&=k_i,\quad i \ne x.
  \end{align}
It then follows that
  \begin{align} \label{Eq: P -vf/2}
     P^{(1)}-\frac{vf}{2}=\int_{k_x'>0} \!\!\!\! {{\dbar}\bk'_\|} \int_{-v k'_x/2}^{v k'_x/2}{\dbar} \omega'\hbar \omega' \,\frac{e^{-2 |k_\perp| d}\, 2 \im R^{-}_{\omega',\bk'_\|} \, 2 \im R^{+}_{\omega',\bk'_\|}}{{|1- e^{2i k_\perp d}R^{-}_{\omega',\bk'_\|} R^{+}_{\omega',\bk'_\|}|^2}} \,,
  \end{align}
where $R^{+}$ and $R^{-}$ are the reflection matrices from half-spaces moving at velocities $v/2$ and $-v/2$, respectively, along the $x$ axis. Since $\epsilon$ is real (the real part of the response function is even in frequency, i.e. $\re \epsilon(\omega)=\re \epsilon(-\omega)$), we have
  \begin{equation}
    R^+_{-\omega',\bk'_\|}=    R^-_{\omega',\bk'_\|}.
  \end{equation}
  This implies that the integrand in Eq.~(\ref{Eq: P -vf/2}) is antisymmetric with respect to $\omega'$ so that the integral vanishes.

    When there is friction, work must be done to keep the moving half-space in steady motion. This work should be equal to the total energy dissipated in the half-spaces,
  \begin{equation}\label{Eq: energy cons}
    v f= P_{\rm tot},
  \end{equation}
  where $P_{\rm tot}$ is the sum of energy flux through each half-space. For Eq.~(\ref{Eq: energy cons}) to hold, the energy flux through the second (moving) half-space should also be equal to $P^{(1)}=vf/2$. In the \emph{center-of-mass }frame too, we should have the same condition because of the energy conservation $vf= P_{\rm c.m.}^{(1)}+P_{\rm c.m.}^{(2)}$, and the symmetry $P_{\rm c.m.}^{(1)}=P_{\rm c.m.}^{(2)}$. (The force in the \emph{center-of-mass} frame is almost identical to the lab frame since the velocity is small compared to the speed of light.)
  Therefore, we conjecture that $P_{S}^{(1)}=P_{S}^{(2)}=vf/2$ irrespective of the reference frame $S$, while $P^{S}_{\rm gap}$ sensitively depends on the reference frame $S$; it is $vf/2$ when the first half-space is at rest, $-vf/2$ when the second half-space is at rest, and zero in the \emph{center of mass} frame.

\section{Discussion and Summary}
Throughout this paper, we explicitly considered half-spaces described by a constant and real dielectric function. However, the underlying physics is rather general, and does not depend on the idealizations made for the sake of convenience.
For example, rather than a half-space, we can consider a thick slab of a material with a complex dielectric function $\epsilon$.
The slab will act like an infinite medium provided that the imaginary part of
$\epsilon$ while small, is sufficiently large to absorb the emitted energy within
the slab, with almost no radiation escaping the far end.
Such conditions can be met for a broad range of the thickness and lossy-ness.

Classically, Cherenkov radiation is emitted when a charged particle passes through a medium. However, even a source {\it without} a net charge, or even a multipole moment, may result in Cherenkov radiation due to quantum fluctuations~\cite{Frank45}.
In the present paper, this is demonstrated for two neutral parallel half-spaces in relative motion.
By employing an amalgam of techniques, usually applied in different contexts,
we are able to make novel conceptual and technical observations.
These techniques are applicable to a variety of other setups:
An interesting situation, closer in spirit to Cherenkov radiation, is when a particle passes through a small channel drilled into a dielectric. Another closely related problem is a particle moving parallel to a surface \cite{Barton10,zhao12,Maghrebi13}. A classical analogue of the latter, namely a charged particle moving above a dielectric half-space, is discussed in Ref.~\cite{Schieber98}.
Our approach of utilizing scattering theory in conjunction with a host of other methods, including input-output and Rytov formalism, should be useful in analyzing such situations.

\section*{Acknowledgements}
    This work is supported by the U.S. Department of Energy under cooperative research agreement Contract Number DE-FG02-05ER41360 (MFM), and the National Science Foundation under Grants No. DMR12-06323 (MK) and NSF PHY11-25915 (RG and MK).

\appendix
\numberwithin{equation}{section}
\section{Green's functions}
In this appendix we compute a number of Green's functions where the two spatial arguments lie within or outside each half-space. We take the first half-space to be at rest while the second one is moving at a velocity $\bv$ parallel to its surface. We further assume that $|\bv|\ll c$ for simplicity.

\noindent  $\bullet \,$ {\it Green's function with both points lying within the gap (outside both objects)}: In this case, the Green's function is given by (with $z_\bx>z_\bz$)
  \begin{align} \label{Eq: G out-out}
    G_{\rm out-out}(\omega,\bx,\bz)= \sum_{\alpha} \frac{1}{2 i p_\alpha} \frac{e^{i p_\alpha d}}{1- e^{2i p_\alpha d}R_\alpha \tilde R_\alpha} &(\Phi^{\reg}_\bbalpha(\omega, \tilde\bx) +\tilde R_\bbalpha \Phi_\bbalpha^{\out}(\omega,\tilde \bx)) \times \nonumber \\
    & (\Phi_\alpha^{\reg}(\omega,\bz)+R_\alpha \Phi_{\alpha}^{\out}(\omega,\bz)),
  \end{align}
  where we used a compact notation defined as
  \begin{align}
  &\alpha=\bk_\| \,, \quad
      \bbalpha=-\bk_\| \,, \quad
      p_\alpha= k_\perp=\sqrt{\omega^2-\bk_\|^2} \,,         \cr
  &    \Phi_\alpha^{\reg}=e^{i\bk_\|\cdot \bx_\|+i k_\perp z } \,, \qquad
      \Phi_{\alpha}^{\out}=e^{i\bk_\|\cdot \bx_\|-i k_\perp z } \,, \cr
   &   R_{\alpha}\equiv R_1(\omega, \bk_\|)=R_{\omega \bk_\|} \,, \qquad
      \tilde R_{\alpha}\equiv R_2(\omega, \bk_\|)=R_{\omega' \bk'_\|} \,, \cr
    &  z_{\tilde \bx}=d-z_{\bx}, \qquad \tilde \bx_\|= \bx_\|, \qquad \sum_\alpha=L^{D-1}\int {\dbar} \bk_\| \equiv L^{D-1}\int \frac{d^{D-1}\bk_\|}{(2\pi)^{D-1}} \,,
  \end{align}
  where $\omega'$ and $\bk_\|'$ are the Lorentz transformation of $\omega$ and $\bk_\|$, respectively. Also $D$ is the number of (spatial) dimensions. According to these definitions, $\Phi_\alpha(\bz)$ is defined with respect to an origin on the surface of the first half-space, while $\Phi_\alpha(\tilde \bx)$ is the wavefunction defined with the origin on the surface of the second half-space and the direction of the $z$-axis reversed. It is straightforward to check that the expression in Eq.~(\ref{Eq: G out-out}) is indeed the Green's function. First note that, for $\bx\ne \bz$, it solves the homogenous version of Eq.~(\ref{Eq: Green's function}). Furthermore, the coefficients are chosen to produce a delta function when $\bz \to \bx$ upon applying the Helmholtz operator.

\noindent  $\bullet \,$ {\it Green's function with one point in the gap and the other inside the first half-space}: This Green's function can be obtained from continuity conditions, {i.e.} by matching the Green's functions approaching a point on the boundary from inside and outside the object
  \begin{equation}
    G_{\rm out-out}(\omega, \bx,\by){\mid}_{\by\to \Sigma^{+}}=  G_{\rm out-in}(\omega, \bx,\by){\mid}_{\by \to \Sigma^-}.
  \end{equation}
This leads to
  \begin{align} \label{Eq: G out-in}
   \!\!\!\!\!\!\! G_{\rm out-in}(\omega,\bx,\bz)= \sum_{\alpha} 1/(2 i p_\alpha) \frac{e^{i p_\alpha d}}{1- e^{2i p_\alpha d}R_\alpha \tilde R_\alpha} & (\Phi^{\reg}_\bbalpha(\omega, \tilde\bx) +\tilde R_\bbalpha \Phi_\bbalpha^{\out}(\omega,\tilde \bx)) \times \nonumber \\
    & (V_{\alpha}\psi_\alpha^{+}(\omega,\bz)+W_\alpha \psi_{\alpha}^{-}(\omega,\bz))\,,
  \end{align}
  with
  \begin{align}
       \psi_{\alpha}^{\pm}=e^{i\bk_\|\cdot \bx_\|\pm i \tilde k_\perp z },
  \end{align}
  where $\tilde k_\perp\equiv \tilde p_\alpha=\sqrt{\epsilon \,\omega^2/c^2-\bk_\|^2}$.
  The (diagonal) matrices $V$ and $W$ are determined by imposing continuity equations, as
  \begin{align}
      V_\alpha +W_\alpha &=1 +R_\alpha \,, \cr
       \tilde p_\alpha (V_\alpha -W_\alpha)&=p_\alpha(1-R_\alpha)\,.
  \end{align}

\noindent  $\bullet \,$ {\it Green's function with both points inside the first half-space}: This Green's function is given by ($z_\bx>z_\bz$)
  \begin{align}\label{Eq: G in-in}
    \!\!\!\!\!\!\!\!\!\!\!\!\!\!\!\! G_{\rm in-in}(\omega,\bx,\bz)= \sum_{\alpha} 1/(2 i p_\alpha) \frac{e^{i p_\alpha d}}{1- e^{2i p_\alpha d}R_\alpha \tilde R_\alpha}     &(V_{\alpha}\psi_\alpha^{+}(\omega,\bx)+W_\alpha \psi_{\alpha}^{-}(\omega,\bx))
    \times \nonumber \\
    &(\tilde V_{\bbalpha}\psi_\bbalpha^{+}(\omega,\bz)+\tilde W_\bbalpha \psi_{\bbalpha}^{-}(\omega,\bz))\,,
  \end{align}
  where, via continuity relations, we have
  \begin{align}
      \tilde V_\alpha +\tilde W_\alpha &=e^{i p_\alpha d}(1 +\tilde R_\alpha) \,, \cr
       \tilde p_\alpha (\tilde V_\alpha -\tilde W_\alpha)&=p_\alpha e^{i p_\alpha d} (1-\tilde R_\alpha).
  \end{align}


\begin{thebibliography}{34}
\expandafter\ifx\csname natexlab\endcsname\relax\def\natexlab#1{#1}\fi
\expandafter\ifx\csname bibnamefont\endcsname\relax
  \def\bibnamefont#1{#1}\fi
\expandafter\ifx\csname bibfnamefont\endcsname\relax
  \def\bibfnamefont#1{#1}\fi
\expandafter\ifx\csname citenamefont\endcsname\relax
  \def\citenamefont#1{#1}\fi
\expandafter\ifx\csname url\endcsname\relax
  \def\url#1{\texttt{#1}}\fi
\expandafter\ifx\csname urlprefix\endcsname\relax\def\urlprefix{URL }\fi
\providecommand{\bibinfo}[2]{#2}
\providecommand{\eprint}[2][]{\url{#2}}

\bibitem[{\citenamefont{Pendry}(1997)}]{Pendry97}
\bibinfo{author}{\bibfnamefont{J.~B.} \bibnamefont{Pendry}},
  \bibinfo{journal}{J. Phys.-Condens. Mat.} \textbf{\bibinfo{volume}{9}},
  \bibinfo{pages}{10301} (\bibinfo{year}{1997}).

\bibitem[{\citenamefont{Volokitin and Persson}(1999)}]{Volokitin99}
\bibinfo{author}{\bibfnamefont{A.~I.} \bibnamefont{Volokitin}}
  \bibnamefont{and} \bibinfo{author}{\bibfnamefont{B.~N.~J.}
  \bibnamefont{Persson}}, \bibinfo{journal}{J. Phys.-Condens. Mat.}
  \textbf{\bibinfo{volume}{11}}, \bibinfo{pages}{345} (\bibinfo{year}{1999}).

\bibitem[{\citenamefont{Volokitin and Persson}(2007)}]{Volokitin07}
\bibinfo{author}{\bibfnamefont{A.~I.} \bibnamefont{Volokitin}}
  \bibnamefont{and} \bibinfo{author}{\bibfnamefont{B.~N.~J.}
  \bibnamefont{Persson}}, \bibinfo{journal}{Rev. Mod. Phys.}
  \textbf{\bibinfo{volume}{79}}, \bibinfo{pages}{1291} (\bibinfo{year}{2007}).

\bibitem[{\citenamefont{{Dalvit} et~al.}(2011)\citenamefont{{Dalvit}, {Maia
  Neto}, and {Mazzitelli}}}]{Dalvit11}
\bibinfo{author}{\bibfnamefont{D.~A.~R.} \bibnamefont{{Dalvit}}},
  \bibinfo{author}{\bibfnamefont{P.~A.} \bibnamefont{{Maia Neto}}},
  \bibnamefont{and} \bibinfo{author}{\bibfnamefont{F.~D.}
  \bibnamefont{{Mazzitelli}}}, in \emph{\bibinfo{booktitle}{Lecture Notes in
  Physics}} (\bibinfo{year}{2011}), vol. \bibinfo{volume}{834}, p.
  \bibinfo{pages}{419}.

\bibitem[{\citenamefont{Frank and Ginzburg}(1945)}]{Frank45}
\bibinfo{author}{\bibfnamefont{I.~M.} \bibnamefont{Frank}} \bibnamefont{and}
  \bibinfo{author}{\bibfnamefont{V.~L.} \bibnamefont{Ginzburg}},
  \bibinfo{journal}{J. Phys. (USSR)} \textbf{\bibinfo{volume}{9}},
  \bibinfo{pages}{353} (\bibinfo{year}{1945}).

\bibitem[{\citenamefont{Ginzburg}(1993)}]{Ginzburg93}
\bibinfo{author}{\bibfnamefont{V.~L.} \bibnamefont{Ginzburg}}
  (\bibinfo{publisher}{Elsevier}, \bibinfo{year}{1993}),
  vol.~\bibinfo{volume}{32} of \emph{\bibinfo{series}{Progress in Optics}}, pp.
  \bibinfo{pages}{267 -- 312}.

\bibitem[{\citenamefont{Ginzburg}(1996)}]{Ginzburg96}
\bibinfo{author}{\bibfnamefont{V.~L.} \bibnamefont{Ginzburg}},
  \bibinfo{journal}{Physics-Uspekhi} \textbf{\bibinfo{volume}{39}},
  \bibinfo{pages}{973} (\bibinfo{year}{1996}).

\bibitem[{\citenamefont{Zel'dovich}(1971)}]{Zel'dovich71}
\bibinfo{author}{\bibfnamefont{Y.~B.} \bibnamefont{Zel'dovich}},
  \bibinfo{journal}{JETP Lett.} \textbf{\bibinfo{volume}{14}},
  \bibinfo{pages}{180} (\bibinfo{year}{1971}).

\bibitem[{\citenamefont{Bekenstein and Schiffer}(1998)}]{Bekenstein98}
\bibinfo{author}{\bibfnamefont{J.~D.} \bibnamefont{Bekenstein}}
  \bibnamefont{and} \bibinfo{author}{\bibfnamefont{M.}~\bibnamefont{Schiffer}},
  \bibinfo{journal}{Phys. Rev. D} \textbf{\bibinfo{volume}{58}},
  \bibinfo{pages}{064014} (\bibinfo{year}{1998}).

\bibitem[{\citenamefont{{Rytov} et~al.}(1989)\citenamefont{{Rytov}, {Kravtsov},
  and {Tatarskii}}}]{Rytov89}
\bibinfo{author}{\bibfnamefont{S.~M.} \bibnamefont{{Rytov}}},
  \bibinfo{author}{\bibfnamefont{Y.~A.} \bibnamefont{{Kravtsov}}},
  \bibnamefont{and} \bibinfo{author}{\bibfnamefont{V.~I.}
  \bibnamefont{{Tatarskii}}}, \emph{\bibinfo{title}{{Priniciples of statistical
  radiophysics. 3. Elements of random fields.}}}, vol.~\bibinfo{volume}{3}
  (\bibinfo{publisher}{Springer}, \bibinfo{address}{Berlin},
  \bibinfo{year}{1989}).

\bibitem[{\citenamefont{Dedkov and Kyasov}(2010)}]{dedkov10}
\bibinfo{author}{\bibfnamefont{G.}~\bibnamefont{Dedkov}} \bibnamefont{and}
  \bibinfo{author}{\bibfnamefont{A.}~\bibnamefont{Kyasov}},
  \bibinfo{journal}{Surface Science} \textbf{\bibinfo{volume}{604}},
  \bibinfo{pages}{562} (\bibinfo{year}{2010}).

\bibitem[{\citenamefont{Barton}(2011)}]{barton11}
\bibinfo{author}{\bibfnamefont{G.}~\bibnamefont{Barton}},
  \bibinfo{journal}{Journal of Physics: Condensed Matter}
  \textbf{\bibinfo{volume}{23}}, \bibinfo{pages}{355004}
  (\bibinfo{year}{2011}).

\bibitem[{\citenamefont{Landau}(1941)}]{Landau41}
\bibinfo{author}{\bibfnamefont{L.}~\bibnamefont{Landau}}, \bibinfo{journal}{Zh.
  Eksp. Teor. Fiz.} \textbf{\bibinfo{volume}{11}}, \bibinfo{pages}{592}
  (\bibinfo{year}{1941}).

\bibitem[{\citenamefont{Matloob et~al.}(1995)\citenamefont{Matloob, Loudon,
  Barnett, and Jeffers}}]{Matloob95}
\bibinfo{author}{\bibfnamefont{R.}~\bibnamefont{Matloob}},
  \bibinfo{author}{\bibfnamefont{R.}~\bibnamefont{Loudon}},
  \bibinfo{author}{\bibfnamefont{S.~M.} \bibnamefont{Barnett}},
  \bibnamefont{and} \bibinfo{author}{\bibfnamefont{J.}~\bibnamefont{Jeffers}},
  \bibinfo{journal}{Phys. Rev. A} \textbf{\bibinfo{volume}{52}},
  \bibinfo{pages}{4823} (\bibinfo{year}{1995}).

\bibitem[{\citenamefont{Gruner and Welsch}(1996)}]{Gruner96}
\bibinfo{author}{\bibfnamefont{T.}~\bibnamefont{Gruner}} \bibnamefont{and}
  \bibinfo{author}{\bibfnamefont{D.-G.} \bibnamefont{Welsch}},
  \bibinfo{journal}{Phys. Rev. A} \textbf{\bibinfo{volume}{54}},
  \bibinfo{pages}{1661} (\bibinfo{year}{1996}).

\bibitem[{\citenamefont{Artoni and Loudon}(1997)}]{Loudon97}
\bibinfo{author}{\bibfnamefont{M.}~\bibnamefont{Artoni}} \bibnamefont{and}
  \bibinfo{author}{\bibfnamefont{R.}~\bibnamefont{Loudon}},
  \bibinfo{journal}{Phys. Rev. A} \textbf{\bibinfo{volume}{55}},
  \bibinfo{pages}{1347} (\bibinfo{year}{1997}).

\bibitem[{\citenamefont{Beenakker}(1998)}]{Beenakker98}
\bibinfo{author}{\bibfnamefont{C.~W.~J.} \bibnamefont{Beenakker}},
  \bibinfo{journal}{Phys. Rev. Lett.} \textbf{\bibinfo{volume}{81}},
  \bibinfo{pages}{1829} (\bibinfo{year}{1998}).

\bibitem[{\citenamefont{{Fulling} and {Davies}}(1976)}]{Fulling76}
\bibinfo{author}{\bibfnamefont{S.~A.} \bibnamefont{{Fulling}}}
  \bibnamefont{and} \bibinfo{author}{\bibfnamefont{P.~C.~W.}
  \bibnamefont{{Davies}}}, \bibinfo{journal}{P. Roy. Soc. Lond. A Mat.}
  \textbf{\bibinfo{volume}{348}}, \bibinfo{pages}{393} (\bibinfo{year}{1976}).

\bibitem[{\citenamefont{Neto and Machado}(1996)}]{Neto96}
\bibinfo{author}{\bibfnamefont{P.~A.~M.} \bibnamefont{Neto}} \bibnamefont{and}
  \bibinfo{author}{\bibfnamefont{L.~A.~S.} \bibnamefont{Machado}},
  \bibinfo{journal}{Phys. Rev. A} \textbf{\bibinfo{volume}{54}},
  \bibinfo{pages}{3420} (\bibinfo{year}{1996}).

\bibitem[{\citenamefont{Lambrecht et~al.}(1996)\citenamefont{Lambrecht, Jaekel,
  and Reynaud}}]{Lambrecht96}
\bibinfo{author}{\bibfnamefont{A.}~\bibnamefont{Lambrecht}},
  \bibinfo{author}{\bibfnamefont{M.-T.} \bibnamefont{Jaekel}},
  \bibnamefont{and} \bibinfo{author}{\bibfnamefont{S.}~\bibnamefont{Reynaud}},
  \bibinfo{journal}{Phys. Rev. Lett.} \textbf{\bibinfo{volume}{77}},
  \bibinfo{pages}{615} (\bibinfo{year}{1996}).

\bibitem[{\citenamefont{{Wilson} et~al.}(2011)\citenamefont{{Wilson},
  {Johansson}, {Pourkabirian}, {Simoen}, {Johansson}, {Duty}, {Nori}, and
  {Delsing}}}]{Nori11}
\bibinfo{author}{\bibfnamefont{C.~M.} \bibnamefont{{Wilson}}},
  \bibinfo{author}{\bibfnamefont{G.}~\bibnamefont{{Johansson}}},
  \bibinfo{author}{\bibfnamefont{A.}~\bibnamefont{{Pourkabirian}}},
  \bibinfo{author}{\bibfnamefont{M.}~\bibnamefont{{Simoen}}},
  \bibinfo{author}{\bibfnamefont{J.~R.} \bibnamefont{{Johansson}}},
  \bibinfo{author}{\bibfnamefont{T.}~\bibnamefont{{Duty}}},
  \bibinfo{author}{\bibfnamefont{F.}~\bibnamefont{{Nori}}}, \bibnamefont{and}
  \bibinfo{author}{\bibfnamefont{P.}~\bibnamefont{{Delsing}}},
  \bibinfo{journal}{nature} \textbf{\bibinfo{volume}{479}},
  \bibinfo{pages}{376} (\bibinfo{year}{2011}).

\bibitem[{\citenamefont{Maghrebi et~al.}(2013)\citenamefont{Maghrebi,
  Golestanian, and Kardar}}]{Maghrebi13}
\bibinfo{author}{\bibfnamefont{M.~F.} \bibnamefont{Maghrebi}},
  \bibinfo{author}{\bibfnamefont{R.}~\bibnamefont{Golestanian}},
  \bibnamefont{and} \bibinfo{author}{\bibfnamefont{M.}~\bibnamefont{Kardar}},
  \bibinfo{journal}{Phys. Rev. D} \textbf{\bibinfo{volume}{87}},
  \bibinfo{pages}{025016} (\bibinfo{year}{2013}).

\bibitem[{\citenamefont{{Penrose}}(1969)}]{Penrose69}
\bibinfo{author}{\bibfnamefont{R.}~\bibnamefont{{Penrose}}},
  \bibinfo{journal}{Nuovo Cimento Rivista Serie} \textbf{\bibinfo{volume}{1}},
  \bibinfo{pages}{252} (\bibinfo{year}{1969}).

\bibitem[{\citenamefont{Maghrebi et~al.}(2012)\citenamefont{Maghrebi, Jaffe,
  and Kardar}}]{Maghrebi12}
\bibinfo{author}{\bibfnamefont{M.~F.} \bibnamefont{Maghrebi}},
  \bibinfo{author}{\bibfnamefont{R.~L.} \bibnamefont{Jaffe}}, \bibnamefont{and}
  \bibinfo{author}{\bibfnamefont{M.}~\bibnamefont{Kardar}},
  \bibinfo{journal}{Phys. Rev. Lett.} \textbf{\bibinfo{volume}{108}},
  \bibinfo{pages}{230403} (\bibinfo{year}{2012}).

\bibitem[{\citenamefont{Manjavacas and Garc\'{\i}a~de
  Abajo}(2010)}]{Manjavacas10}
\bibinfo{author}{\bibfnamefont{A.}~\bibnamefont{Manjavacas}} \bibnamefont{and}
  \bibinfo{author}{\bibfnamefont{F.~J.} \bibnamefont{Garc\'{\i}a~de Abajo}},
  \bibinfo{journal}{Phys. Rev. Lett.} \textbf{\bibinfo{volume}{105}},
  \bibinfo{pages}{113601} (\bibinfo{year}{2010}).

\bibitem[{\citenamefont{Unruh}(1974)}]{Unruh74}
\bibinfo{author}{\bibfnamefont{W.~G.} \bibnamefont{Unruh}},
  \bibinfo{journal}{Phys. Rev. D} \textbf{\bibinfo{volume}{10}},
  \bibinfo{pages}{3194} (\bibinfo{year}{1974}).

\bibitem[{\citenamefont{Birrell and Davies}(1984)}]{birrell1984}
\bibinfo{author}{\bibfnamefont{N.~D.} \bibnamefont{Birrell}} \bibnamefont{and}
  \bibinfo{author}{\bibfnamefont{P.~C.~W.} \bibnamefont{Davies}},
  \emph{\bibinfo{title}{Quantum fields in curved space}}, \bibinfo{number}{7}
  (\bibinfo{publisher}{Cambridge University Press}, \bibinfo{year}{1984}).

\bibitem[{\citenamefont{Fulling}(1989)}]{fulling1989}
\bibinfo{author}{\bibfnamefont{S.~A.} \bibnamefont{Fulling}},
  \emph{\bibinfo{title}{Aspects of quantum field theory in curved spacetime}},
  vol.~\bibinfo{volume}{17} (\bibinfo{publisher}{Cambridge University Press},
  \bibinfo{year}{1989}).

\bibitem[{\citenamefont{{Carroll}}(2004)}]{Carroll04}
\bibinfo{author}{\bibfnamefont{S.~M.} \bibnamefont{{Carroll}}},
  \emph{\bibinfo{title}{{Spacetime and geometry. An introduction to general
  relativity}}} (\bibinfo{publisher}{Addison Wesley}, \bibinfo{year}{2004}).

\bibitem[{\citenamefont{Kr\"uger et~al.}(2011)\citenamefont{Kr\"uger, Emig, and
  Kardar}}]{Kruger11}
\bibinfo{author}{\bibfnamefont{M.}~\bibnamefont{Kr\"uger}},
  \bibinfo{author}{\bibfnamefont{T.}~\bibnamefont{Emig}}, \bibnamefont{and}
  \bibinfo{author}{\bibfnamefont{M.}~\bibnamefont{Kardar}},
  \bibinfo{journal}{Phys. Rev. Lett.} \textbf{\bibinfo{volume}{106}},
  \bibinfo{pages}{210404} (\bibinfo{year}{2011}).

\bibitem[{\citenamefont{Jackson}(1998)}]{Jackson98}
\bibinfo{author}{\bibfnamefont{J.~D.} \bibnamefont{Jackson}},
  \emph{\bibinfo{title}{Classical Electrodynamics}}
  (\bibinfo{publisher}{Wiley}, \bibinfo{address}{New York},
  \bibinfo{year}{1998}), \bibinfo{edition}{3rd} ed.

\bibitem[{\citenamefont{Barton}(2010)}]{Barton10}
\bibinfo{author}{\bibfnamefont{G.}~\bibnamefont{Barton}}, \bibinfo{journal}{New
  J. Phys.} \textbf{\bibinfo{volume}{12}}, \bibinfo{pages}{113045}
  (\bibinfo{year}{2010}).

\bibitem[{\citenamefont{Zhao et~al.}(2012)\citenamefont{Zhao, Manjavacas,
  Garc\'{\i}a~de Abajo, and Pendry}}]{zhao12}
\bibinfo{author}{\bibfnamefont{R.}~\bibnamefont{Zhao}},
  \bibinfo{author}{\bibfnamefont{A.}~\bibnamefont{Manjavacas}},
  \bibinfo{author}{\bibfnamefont{F.~J.} \bibnamefont{Garc\'{\i}a~de Abajo}},
  \bibnamefont{and} \bibinfo{author}{\bibfnamefont{J.~B.}
  \bibnamefont{Pendry}}, \bibinfo{journal}{Phys. Rev. Lett.}
  \textbf{\bibinfo{volume}{109}}, \bibinfo{pages}{123604}
  (\bibinfo{year}{2012}).

\bibitem[{\citenamefont{Schieber and Sch\"achter}(1998)}]{Schieber98}
\bibinfo{author}{\bibfnamefont{D.}~\bibnamefont{Schieber}} \bibnamefont{and}
  \bibinfo{author}{\bibfnamefont{L.}~\bibnamefont{Sch\"achter}},
  \bibinfo{journal}{Phys. Rev. E} \textbf{\bibinfo{volume}{57}},
  \bibinfo{pages}{6008} (\bibinfo{year}{1998}).

\end{thebibliography}
\end{document}